\documentstyle[prl,aps,multicol,epsf]{revtex}
 
\input{epsf} 
\epsfverbosetrue
\epsfclipon

\newcommand{\putfigxsz}[4]{
   \begin{figure}\begin{center}\mbox{\epsfxsize #4
   \epsffile{#1}}
   \end{center}
   \end{figure}
   }

\begin{document}
\draft
\title{Metal-insulator transition at $B$=0
 in an ultra-low density ($r_{\rm{s}}{\simeq}23$)
two dimensional GaAs-AlGaAs hole gas}

\author{M.Y. Simmons$^{(a)}$, A.R. Hamilton$^{(a)}$,
T.G. Griffiths$^{(b)}$, A.K. Savchenko$^{(b)}$,\\ M. Pepper$^{(a)}$,
and D.A. Ritchie$^{(a)}$}

\date{Submitted to EP2DS-12: 21 August 1997}
\address{(a) Cavendish Laboratory, University of Cambridge,
Madingley Road, Cambridge, CB3 OHE, United Kingdom\\
(b) Department of Physics, University of Exeter, Stocker Road,
Exeter EX4 4QL, United Kingdom \vspace{12pt}}
\maketitle

\begin{abstract} 

We have observed a metal-insulator transition in an ultra-low density
two dimensional hole gas formed in a high quality GaAs-AlGaAs
heterostructure at $B=0$. At the highest carrier density studied
($p_{\rm{s}}{=}2.2{\times}10^{10}~\text{cm}^{\text{-2}},
r_{\rm{s}}{=}16$) the hole gas is strongly metallic, with an
exceptional mobility of
$425,000~\text{cm}^2\text{V}^{\text{-1}}\text{s}^{\text{-1}}$.  The
low disorder and strength of the many-body interactions in this sample
are highlighted by the observation of re-entrant metal insulator
transitions in both the fractional ($\nu{<}1/3$) and integer
($2{>}\nu{>}1$) quantum Hall regimes. On reducing the carrier density
the temperature and electric field dependence of the resistivity show
that the sample is still metallic at
$p_{\rm{s}}{=}1.3{\times}10^{10}~\text{cm}^{\text{-2}}$
($r_{\rm{s}}=21$), becoming insulating at
$p_{\rm{s}}{\simeq}1{\times}10^{10}~\text{cm}^{\text{-2}}$. Our
results indicate that electron-electron interactions are dominant at
these low densities, pointing to the many body origins of this
metal-insulator transition. We note that the value of $r_{\rm{s}}$ at
the transition ($r_{\rm{s}}{=}23{\pm}2$) is large enough to allow the
formation of a weakly pinned Wigner crystal, and is approaching the
value calculated for the condensation of a pure Wigner crystal.

\end{abstract}

\begin{multicols}{2}
The scaling theory of localisation\cite{Scaling theory} predicts that
there is no metallic phase in two dimensional systems at zero magnetic
field as $T{\rightarrow}0$, and therefore that there can be no
metal-insulator transition (MIT), in agreement with early experiments
on low mobility silicon MOSFET's~\cite{Pepper}. At low temperatures a
dilute two dimensional system is therefore always insulating and
various models exist to describe the nature of this insulating
state. These range from the single particle Anderson insulator in the
case of strongly disordered systems~\cite{Ando} to the formation of a
pinned Wigner crystal for clean systems with very low
disorder~\cite{Chui}.

In contradiction to the expectations of scaling theory however,
several recent experiments on comparatively high mobility electron
gases in silicon MOSFETs have provided evidence of a metal insulator
transition at zero magnetic field~\cite{Kravchenko,Popovic}.  The
nature of this transition and its physical origins are not presently
understood, but it should be noted that in the original scaling theory
electron-electron interactions were not included.  In particular,
electron-electron interactions become important at low densities in
precisely the regime where the MIT is experimentally observed and will
be dominant if the disorder is low enough.

This work provides the first evidence of a MIT at $B=0$ in GaAs/AlGaAs
heterostructures~\cite{SimmonsWeb}. Using a back-gate we study the
transition from a strongly metallic state to an insulating state at
$B=0$ in an extremely high quality, low density two dimensional hole
gas as the carrier density is reduced. The large effective mass
($m^{*} \simeq 0.3m_e$) and the extremely low disorder (with
mobilities an order of magnitude larger than in Si MOSFETs) enhance
interaction effects, the strength of which are characterised by the
dimensionless parameter $r_{\rm{s}}$, the ratio of the Coulomb
interaction energy to the kinetic (Fermi) energy, given by
$r_{\rm{s}}=m^{*}e^{2}/(4\pi\epsilon\hbar^{2}\sqrt{\pi p_{\rm{s}}}$),
where $p_{\rm{s}}$ is the sheet carrier density. The MIT observed in
electron gases in Si MOSFETs occurs at
$r_{\rm{s}}\simeq10$~\cite{Kravchenko,Popovic}; in contrast for our
hole gas samples we observe strongly metallic behaviour at
$r_{\rm{s}}=16$ with a MIT occurring at $r_{\rm{s}}=23\pm 3$. This
value of $r_{\rm{s}}$ is significantly larger than that found for the
MIT observed in Si MOSFETs and is approaching the value of
$r_{\rm{s}}$ at which Wigner crystallisation is expected to occur in a
perfectly clean 2D system ($r_{\rm{s}}=37\pm 5$)~\cite{Chui2}.

The heterostructure used in this study (T129) was fabricated by MBE
growth on the (311)A surface of GaAs, utilising silicon as the
acceptor dopant~\cite{Simmons}. The structure consisted of a 300{\AA}
p-type modulation doped GaAs quantum well with a 1600{\AA}
Al$_{0.33}$Ga$_{0.67}$As spacer and an average distance of 2100{\AA}
between the carriers and the remote impurities.  Samples were
patterned into Hall bars of size $60{\times}600~{\mu}m$ aligned in the
$[\bar233]$ direction. Two different samples were studied from the
same wafer: Sample 1 was measured in the mixing chamber of a dilution
refrigerator; sample 2 had an external metal plate approximately
500~${\mu}m$ away from the 2DHG acting as a back-gate and was measured
on a cold finger. In both cases, magnetotransport measurements were
made in the dark at a frequency of below 10~Hz, with a current of
$<2~n$A (sample 1) or an excitation voltage of
V$_{\rm{ac}}<100~{\mu}$V (sample 2).

Figure~\ref{highnSdH} shows the temperature dependence of the
magnetotransport data for sample 1 at a carrier density of
$p_{\rm{s}}{=}2.2{\times}10^{10}~\text{cm}^{\text{-2}}$
($r_{\rm{s}}$=16).  The exceptionally high quality of this sample is
highlighted by the clear observation of fractional quantum Hall (FQH)
states at $\nu$=2/3, 2/5, and 1/3.  Indeed at the lowest temperatures
studied, and with a current of 0.04~nA a clear $\rho_{xx}$ minima at
$\nu$=2/7 was observed; evidence of this state can also be seen in the
$\rho_{xy}$ data at $B$=3.3~T. Several metal-insulator transitions can
be seen in the data.  Above $B$=1.5~T a weakly insulating state is
observed, which is interrupted by the $\nu$=1/3 FQH liquid, leading
finally to a strongly insulating state at $\nu{<}$1/3.

\begin{minipage}{7.2cm}
\noindent
\begin{figure}[tbph]
\putfigxsz{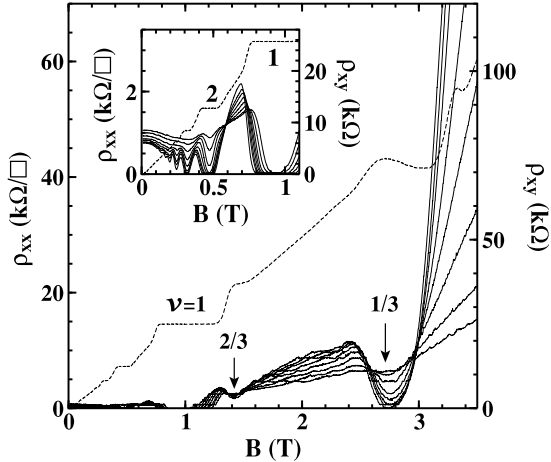}{lab}{cap}{7.2cm}
\caption{Magnetotransport data for sample 1, measured at
T=50, 77, 98, 133, 180, 240, 333 and 425 mK.
The inset shows the low field data.} 
\label{highnSdH} 
\end{figure}                  
\end{minipage}
\vspace{0.1in}

The global phase diagram of Kivelson {\it et al.}~\cite{Kivelson}
predicts a transition from a quantum Hall liquid at $\nu$=1/3 to a
Hall insulator. However, re-entrant insulating behaviour similar to
that seen in figure 1 has been observed around $\nu$=1/3 in higher
density hole gases~\cite{Santos}, and $\nu$=1/5 in electron
systems~\cite{Goldman} and has been taken as evidence for the
formation of a weakly pinned magnetically induced Wigner crystal.

The inset to Fig.~\ref{highnSdH} shows the low field integer quantum
Hall effect for $\nu$=1--6.  A second distinct metal-insulator
transition occurs between $\nu$=1 and 2, with a critical point at
which $\rho_{xx}$ is $T$ independent at $B_{\rm{c}}$=0.58 T
($\nu_{\rm{c}}$=1.6).  Such a transition is expressly forbidden by the
global phase diagram~\cite{Kivelson}, and can only be accounted for by
extremely strong many-body interactions~\cite{KravchenkoPRL1995}. We
note in passing that this is the first time that metal-insulator
transitions have been simultaneously observed both in the integer and
fractional quantum Hall regimes.

We now consider the behaviour of these many body metal insulator
transitions as the carrier density is reduced with the back-gate.
Figure~\ref{lownSdH}(a) shows magnetoresistance data at different
temperatures for a carrier density of
$p_{\rm{s}}{=}1.3{\times}10^{10}~\text{cm}^{\text{-2}}$
($r_{\rm{s}}$=21).  At this low density, the sample is still metallic
at $B$=0
(${\mu}{=}60,000~\text{cm}^2\text{V}^{\text{-1}}\text{s}^{\text{-1}}$)
and $\rho_{xx}{\rightarrow}0$ as $T{\rightarrow}0$ at $\nu{=}1$, but
the relative disorder has increased. Thus the QH-insulator-QH
transition for $2{>}\nu{>}1$ has disappeared, with the magnetically
induced insulating phase now occurring at $B$=0.8 T ($\nu{<}1$).

\begin{minipage}{7.2cm}
\noindent
\begin{figure}[tbph]
\putfigxsz{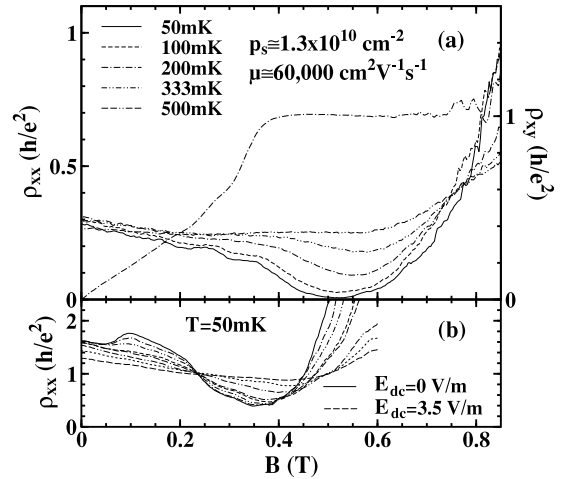}{lab}{cap}{7.2cm}
\caption{Magnetotransport data for sample 2 at low densities, 
(a) at different temperatures
($p_{\rm{s}}{=}1.3{\times}10^{10}~\text{cm}^{\text{-2}}$), and
(b) at electric field values of $E_{\rm{dc}}$ from 0 to 3.5~V/m in
steps of 0.5~V/m along the Hall bar
($p_{\rm{s}}{\simeq}8.5{\times}10^{9}~\text{cm}^{\text{-2}}$).}
\label{lownSdH} 
\end{figure}                  
\end{minipage} 
\vspace{0.1in}

Reducing the carrier density further, to
$p_{\rm{s}}{\simeq}8.5{\times}10^{9}~\text{cm}^{\text{-2}}$, with
$r_{\rm{s}}$=26 (estimated from the minimum in $\rho_{xx}$), causes
the sample to become insulating at $B$=0 as shown in
Fig. \ref{lownSdH}(b). A transition to a $\nu{=}1$ quantum Hall liquid
is observed at $B=0.24$~T. In these measurements we have varied the
electric field $E_{\rm{dc}}$ along the Hall bar rather than varying T,
since the sample was not stable over the large timescales required for
a full temperature dependence.  Nevertheless a clear crossing point is
seen at $B$=0.24 T with a critical resistivity of
$\rho_{xx}^{c}{=}h/e^{2}$.

The metal-insulator transitions shown in Fig.~2 are consistent with
that of a Hall insulator described by Kivelson {\it et
al.}~\cite{Kivelson}. However, it does not follow that the insulating
state observed at $B=0$ is of the same nature. Indeed the MITs
observed in our data are also consistent with the formation of a
pinned Wigner solid. In a perfectly clean system, the formation of a
Wigner solid at $B$=0 is not expected to occur until
$r_{\rm{s}}{\geq}37$. It has been shown, however, that disorder can
stabilise the Wigner solid for $r_{\rm{s}}$ as low as
7.5~\cite{Chui2}. The multiple metal-insulator transitions observed in
Fig.~\ref{highnSdH} highlights the strength of the many body
interactions in our samples and coupled with the large value of
$r_{s}$ suggest that the insulating state observed at zero field
favours that of a Wigner crystal.

Figure~\ref{Tdep}(a) shows the strong linear temperature dependence of
the zero field resistivity for the highest carrier density
($p_{\rm{s}}{=}2.2{\times}10^{10}~\text{cm}^{\text{-2}}$) studied,
characteristic of a metallic state. The change in zero field
resistivity observed is too large to be caused by quantum interference
effects and in the absence of phonon scattering (we are well within
the Bloch-Gr\"{u}neisen regime) we attribute this linear dependence to
the temperature dependence of the screening~\cite{GoldDasSarma}.

\begin{minipage}{7.2cm}
\noindent
\begin{figure}[tbph]
\putfigxsz{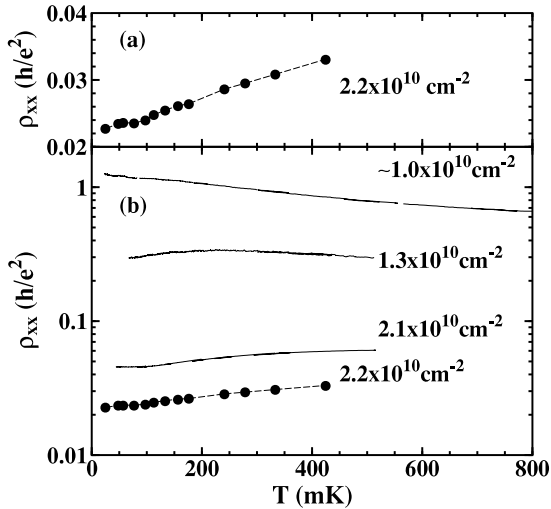}{lab}{cap}{7.2cm}
\caption{(a) T-dependence of $\rho_{xx}$ for sample 1 at
$p_{\rm{s}}{=}2.2{\times}10^{10}~\text{cm}^{\text{-2}}$ on a linear
scale.  (b) T-dependence of $\rho_{xx}$ for sample 2 at different
densities.}
\label{Tdep}  
\end{figure}                  
\end{minipage} 
\vspace{0.1in}

To further examine the nature of the $B$=0 metal-insulator transition
we show the temperature dependence of the zero field resistivity as we
reduce the carrier density in figure~\ref{Tdep}(b). In the
intermediate density range
($p_{s}=1.3{\times}10^{10}~\text{cm}^{\text{-2}}$) we observe a change
in sign of $\partial\rho_{xx}/\partial{T}$, with
$\partial\rho_{xx}/\partial{T}{>}0$ for $T{<}300~$mK, and
$\partial\rho_{xx}/\partial{T}{<}0$ for $T{>}300~$mK. Similar
behaviour near the MIT transition has been observed in Si
MOSFETs~\cite{Kravchenko} and has been partially ascribed to weak
localisation.  Whilst spin-orbit scattering for hole gases in GaAs is
strong, and a combination of weak localisation and
weak-antilocalisation could cause this non-monotonic
behaviour~\cite{Bergmann}, we do not believe this to be the case. In
particular, we do not observe the temperature dependent low field
negative magnetoresistance that is characteristic of weak localisation
in our data, (Fig~\ref{lownSdH}(a)), nor do we observe a change in the
sign of the low field magnetoresistance correction that would result
in a change from weak localisation to weak anti-localisation with
decreasing temperature. We are therefore drawn to the conclusion that
the temperature dependence of $\rho_{xx}$ in this regime arises from
the increasing importance of electron-electron interactions with
decreasing temperature rather than disorder related weak localisation
effects.

Upon further reducing the hole density to
$p_{s}=1{\times}10^{10}~\text{cm}^{\text{-2}}$, $r_{s}=23$ the hole
gas becomes insulating with $\partial\rho_{xx}/\partial{T}{<}0$ for
all $T$. Whilst the value of $\rho_{xx}$ at the transition
($\rho_{xx}{\approx}{h/e^{2}}$) is similar to that reported in MITs
observed in Si MOSFETs ($\rho_{xx}{\simeq}{(2-3)h/e^{2}}$), we do not
observe the strong temperature dependence of $\rho_{xx}$ on either
side of the transition reported in reference~\cite{Kravchenko}.

\begin{minipage}{7.2cm}
\noindent
\begin{figure}[tbph]
\putfigxsz{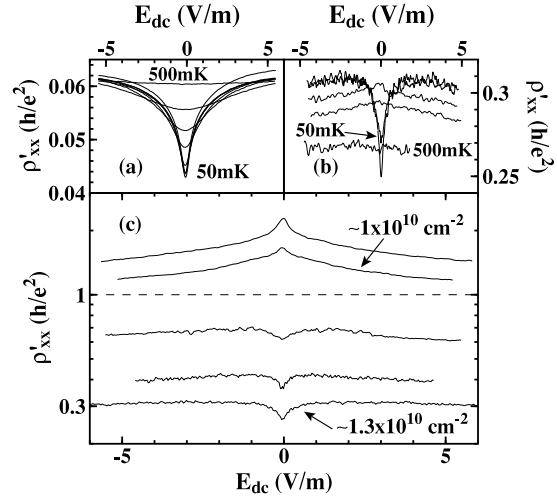}{lab}{cap}{7.2cm}
\caption{(a) The differential resistivity $\rho_{xx}^{\prime}$ as a
function of the electric field $E_{\rm{dc}}$ measured along the
channel, for $T=50$~mK-$500$~mK and
$p_{\rm{s}}{=}2.1{\times}10^{10}~\text{cm}^{\text{-2}}$. (b) Similar
data for $p_{\rm{s}}{=}1.3{\times}10^{10}~\text{cm}^{\text{-2}}$. (c)
$\rho_{xx}^{\prime}(E_{\rm{dc}})$ at $T=30$~mK for several different
carrier densities}
\label{DCbiasing}  \end{figure}                   
\end{minipage}  
\vspace{0.1in}
 
Finally we examine the effects of an electric field on the
differential resistivity on either side of the metal insulator
transition. In Fig.~\ref{DCbiasing} (a) we plot the differential
resistance $\rho_{xx}^{\prime}$ for
$p_{\rm{s}}{=}2.1{\times}10^{10}~\text{cm}^{\text{-2}}$ at
temperatures from $50-500$~mK. At the lowest temperatures there is a
sharp increase in $\rho_{xx}^{\prime}$ with $E_{\rm{dc}}$ consistent
with metallic behaviour, which is washed out as the temperature
increases, with all the curves tending towards the same value of
$\rho_{xx}^{\prime}$ at high $T$ or $E_{\rm{dc}}$.

Fig.~\ref{DCbiasing}(b) shows similar
$\rho_{xx}^{\prime}$($E_{\rm{dc}}$) data for
$p_{\rm{s}}{=}1.3{\times}10^{10}~\text{cm}^{\text{-2}}$. At the lowest
temperatures metallic behaviour is seen, similar to that shown in
Fig.~\ref{DCbiasing}(a). However, as the temperature is increased this
changes to insulating behaviour ($\rho_{xx}^{\prime}$ decreases with
increasing $T$ or $E_{\rm{dc}}$) consistent with the change in sign of
$\partial\rho_{xx}/\partial{T}$ seen in Fig.~\ref{Tdep}(b) at this
density.

The overall trend of $\rho_{xx}^{\prime}$($E_{\rm{dc}}$) for different
carrier densities at a fixed temperature of $T<30$~mK is shown in
Fig.~\ref{DCbiasing}(c), demonstrating the transition from strongly
metallic to insulating behaviour as $p_{s}$ is reduced. The traces are
symmetric about $\rho_{xx}^{c}=h/e^{2}$ (shown by the dashed line in
the figure) in agreement with the value of $\rho_{xx}^{c}$ obtained
from the temperature dependence data of Fig.~\ref{Tdep}(b). Similar
values of $\rho_{xx}^{c}$ and $\rho_{xx}^{\prime}$($E_{\rm{dc}}$) have
recently been reported for high mobility silicon MOSFETs at
$B=0$~\cite{Kravchenko96}. The data in Fig. 4 provides additional
evidence for a true metal insulator transition in high mobility hole
gases in the absence of a magnetic field.

In summary, we have observed a MIT at $B=0$ in a very dilute, low
disorder two dimensional GaAs/AlGaAs hole gas, characterised both by
the $T$ and $E_{\rm{dc}}$ dependence of $\rho_{xx}$. Although the
physical origins of this transition are unclear, many body
interactions are shown to be extremely strong in our samples, with the
value of $r_{\rm{s}}$ large enough to allow the formation of a weakly
pinned Wigner crystal. Our results confirm the universal nature of the
MIT at $B=0$, which has also been observed in different material
systems and with different charge carriers.

This work was funded by the EPSRC. ARH and MYS gratefully acknowledge
financial support from the British Council in Japan; MYS
acknowledges support from the British Association of Crystal Growth;
AKS, TGG, ARH and MYS acknowledge funding from the Exeter
University research fund.

\end{multicols}
\end{document}